\documentclass[a4paper,superscriptaddress,citeautoscript,twocolumn,pra]{revtex4-1}

\usepackage[T1]{fontenc}
\usepackage{times}
\usepackage{amsmath,amssymb}
\usepackage[pdftex]{graphicx,color}
\usepackage{epstopdf}
\usepackage[a4paper,body={18cm,25.5cm},top=2cm]{geometry}
\usepackage[english]{babel}
\usepackage{physics}
\usepackage{orcidlink}

\begin{document}

\title{Dynamics of temporal Kerr cavity solitons in the presence of rapid parameter inhomogeneities: from bichromatic driving to third-order dispersion}

\author{Caleb Todd \orcidlink{0000-0002-3657-8942}}
\altaffiliation{Corresponding author: caleb.todd@auckland.ac.nz}
\author{Zongda Li \orcidlink{0000-0001-8198-0941}}
\affiliation{The Dodd-Walls Centre for Photonic and Quantum Technologies, New Zealand}
\affiliation{Physics Department, The University of Auckland, Private Bag 92019, Auckland 1142, New Zealand}
\author{St\'ephane Coen \orcidlink{0000-0001-5605-5906}}
\affiliation{The Dodd-Walls Centre for Photonic and Quantum Technologies, New Zealand}
\affiliation{Physics Department, The University of Auckland, Private Bag 92019, Auckland 1142, New Zealand}
\author{Stuart G. Murdoch \orcidlink{0000-0002-9169-9472}}
\affiliation{The Dodd-Walls Centre for Photonic and Quantum Technologies, New Zealand}
\affiliation{Physics Department, The University of Auckland, Private Bag 92019, Auckland 1142, New Zealand}
\author{Gian-Luca Oppo \orcidlink{0000-0002-5376-4309}}
\affiliation{SUPA and Department of Physics, University of Strathclyde, Glasgow G4 0NG, Scotland, European Union}
\author{Miro Erkintalo \orcidlink{0000-0001-7753-7007}}
\affiliation{The Dodd-Walls Centre for Photonic and Quantum Technologies, New Zealand}
\affiliation{Physics Department, The University of Auckland, Private Bag 92019, Auckland 1142, New Zealand}

\begin{abstract}
  \noindent Temporal Kerr cavity solitons are pulses of light that can persist in coherently-driven, dispersive resonators with Kerr-type nonlinearity. It is widely accepted that such solitons react to parameter inhomogeneities by experiencing a temporal drift whose rate is governed by the gradient of the parameter at the soliton's position. This result, however, assumes that the gradient of the inhomogeneity is constant across the soliton, which may not hold true under all situations, e.g. when using bichromatic driving or in the presence of third-order dispersion that gives rise to an extended dispersive wave tail. Here we report on theoretical and numerical results pertaining to the behavior of dissipative temporal Kerr cavity solitons under conditions where parameter inhomogeneities vary nonlinearly across the width of the soliton. In this case, the soliton velocity is dictated by the full overlap between its so-called neutral mode and the parameter perturbation, which we show can yield dynamics that are manifestly at odds with the common wisdom of motion dependent solely upon the gradient of the inhomogeneity. We also investigate how the presence of third-order dispersion and the associated dispersive wave tail changes the motion induced by parameter inhomogeneities. We find that the dispersive wave tail as such does not contribute to the soliton motion; instead, higher-order dispersion yields counter-intuitive influences on the soliton motion. Our results provide new insights to the behavior of temporal cavity solitons in the presence of parameter inhomogeneities, and can impact systems employing pulsed or bichromatic pumping and/or resonators with non-negligible higher-order dispersion.

\end{abstract}

\maketitle

\section{Introduction}

Coherently-driven, dispersive resonators with Kerr nonlinearity can sustain localized dissipative structures known as temporal Kerr cavity solitons (TCSs) \cite{wabnitz1993suppression,leo2010temporal, bookcoen2015temporal}. Such solitons, also known as dissipative Kerr solitons, have attracted significant attention over the past decade due to their rich dynamics~\cite{leo_dynamics_2013, anderson_observations_2016, yu_breather_2017, lucas_breathing_2017, cole_soliton_2017, nielsen_nonlinear_2021, xu_spontaneous_2021} as well as significance with regards to practical applications. In particular, TCSs underpin the generation of coherent microresonator optical frequency combs~\cite{coen2013modeling, erkintalo2014coherence, herr2014temporal, kippenberg2018dissipative, pasquazi2018micro}, whose many applications range from telecommunications~\cite{marin2017microresonator} and distance measurements~\cite{suh_soliton_2018, trocha2018ultrafast, riemensberger_massively_2020} to spectroscopy~\cite{suh_microresonator_2016, yu2018silicon} and imaging~\cite{bao2019microresonator}.

At their simplest, TCSs manifest themselves in an environment with full translation invariance. That is to say that the parameters of the system (including the complex amplitude of the coherent field driving the resonator) exhibit negligible variation in time. Under such conditions, the solitons circulate the resonator with a constant (intrinsic) group-velocity, and therefore appear stationary in a retarded temporal reference frame that moves with that same velocity. In stark contrast, the presence of parameter inhomogeneities --- arising e.g. from amplitude or phase modulations applied on the driving field --- can cause the solitons to drift in the solitons' intrinsic reference frame (as defined in the absence of inhomogeneities)~\cite{jang2015temporal, obrzud_temporal_2017, hendry2018spontaneous, cole_Kerr-microresonator_2018, erkintalo2022phase}. The parameter inhomogeneity essentially modifies the solitons' group-velocity, providing valuable means to trap TCSs into dedicated temporal positions or to lock their repetition rate to an external signal~\cite{jang2015temporal, hendry2018spontaneous, brasch_nonlinear_2019, anderson_photonic_2021}.

The motion of \emph{temporal} cavity solitons in the presence of inhomogeneities can be described using theories originally developed in the context of localized structures in \emph{spatial} systems~\cite{firth1996optical, maggipinto2000cavity, scroggie2005reversible, fedorov2001effects}. These theories typically assume that the soliton is infinitely localized --- so the soliton position is uniquely defined --- and that the parameter inhomogeneity varies slowly (i.e., linearly) across its extent; they yield the commonly accepted result that the TCS drift velocity is directly proportional to the gradient of the perturbation. Only a handful of studies (performed in the context of spatially localized structures) have considered the case where the perturbation changes nonlinearly across the soliton~\cite{fedorov2001effects, scroggie2005reversible}, revealing novel dynamics that are at odds with the common wisdom of gradient motion. In dispersive (temporal) systems, the assumption of slow variation of the inhomogeneity is arguably well-justified in many important scenarios, but situations where this is not the case can also be readily envisaged. For instance, the use of bichromatic driving~\cite{hansson_bichromatically_2014,qureshi_soliton_2021,moille_ultra-broadband_2021,taheri2022all} can in principle yield arbitrarily fast (harmonic) temporal inhomogeneities, whilst dispersive waves generated due to higher-order dispersion~\cite{coen2013modeling, milian_soliton_2014, parra2014third, jang_observation_2014} can greatly increase the solitons' temporal extent (and thus render nonlinear parameter variations more likely to occur over that extent). Given the increasing interest in using inhomogeneities to manipulate TCS dynamics across the whole spectrum of applied and fundamental photonics (from frequency comb repetition rate control~\cite{cole_Kerr-microresonator_2018} to the study of dissipative Bloch oscillations~\cite{englebert_bloch_2021}), there is a nascent need to better understand configurations where the inhomogeneity varies nonlinearly across the solitons' extent.

In this paper, we report on a theoretical and numerical study of TCS motion induced by an inhomogeneity that varies nonlinearly across the soliton. We demonstrate that, depending upon the specific parameters, such an inhomogeneity can give rise to soliton motion that is manifestly at odds with simple gradient motion that would be expected for an infinitely-localized soliton subject to a linearly-varying perturbation. We discuss how the motion is dictated (and can thus be predicted) by the overlap between the full parameter perturbation and the so-called neutral mode associated with the soliton~\cite{maggipinto2000cavity, scroggie2005reversible, fedorov2001effects}. We also consider the interplay between parameter inhomogeneities and higher-order dispersion, investigating how the presence of an extended dispersive wave tail affects the motion of TCSs. Surprisingly, we find that the dispersive wave tail as such does not influence the soliton motion; instead, higher-order dispersion can modify the soliton drift dynamics in a highly counter-intuitive fashion: inhomogeneities that do not overlap with the soliton or its dispersive wave can nonetheless induce observable drift. In general, our calculations show that operation close to conditions where higher-order dispersion plays a significant role leads to diminished soliton velocities, thus reducing the effectiveness of using inhomogeneities to control TCSs. Our results shed new light on the inhomogeneity-induced motion of TCSs and could have impact on the control and manipulation of TCSs in the presence of higher-order dispersion or driving fields with rapidly varying phase or amplitude profile.

\section{Theory}

We first recount the basic model and theories of soliton motion pertinent to our study. To this end, we consider a dispersive ring resonator with Kerr-type nonlinearity that is coherently driven with laser light. We focus our attention to the specific scenario where soliton motion is induced by a rapid inhomogeneity associated with the coherent driving field, with the inhomogeneity temporally synchronized with the intrinsic round trip time of the soliton. (We note, however, that our analysis can be readily generalized to account for desynchronisation or inhomogeneities along other system parameters.) Considering dispersion to third-order, the evolution of the slowly-varying electric field envelope $E(t,\tau)$ can be modelled using the following generalized Lugiato-Lefever equation (LLE)~\cite{coen2013modeling}:
\begin{align} \label{lle}
    \pdv{E}{t} = \bigg[ -1 &+ i ( |E|^2 - \Delta ) - d_1 \pdv{}{\tau} + i \pdv[2]{}{\tau} + d_3 \pdv[3]{}{\tau} \bigg] E \nonumber\\[5pt]
    &+ S(\tau).
\end{align}
Here, $t$ is the \emph{slow} time that describes how the intracavity envelope $E(t,\tau)$ changes from roundtrip to roundtrip and $\tau$ is the \emph{fast} time that describes the field profile within a single roundtrip. The terms on the right hand side of Eq.~(\ref{lle}) describe linear losses, Kerr nonlinearity, linear phase detuning of the slowly-varying envelope's carrier frequency from the nearest cavity resonance ($\Delta$ is the detuning parameter), \mbox{first-,} \mbox{second-,} and third-order dispersion, and coherent driving [$S(\tau)$ is the driving field amplitude], respectively. For the normalization of Eq.~(\ref{lle}), see \cite{leo2010temporal}.

The first-order dispersion term (proportional to $d_1$) is included in Eq.~(\ref{lle}) to eliminate soliton drifts that can occur in the absence of parameter inhomogeneities; specifically, the term is used to set the reference frame of Eq.~(\ref{lle}) to be such that it moves with the solitons' intrinsic group-velocity. This choice of reference frame reflects our focus on pump inhomogeneities that are synchronous with the solitons' intrinsic round trip time (note in this context that the driving term $S(\tau)$ is stationary in the sense that it does not depend on the slow time $t$). In the absence of third-order dispersion ($d_3 = 0$), TCSs move with the group velocity of the coherent driving field such that $d_1 = 0$. In contrast, in the presence of third-order dispersion ($d_3 \neq 0$), the solitons experience a group velocity that is different from the group velocity at the carrier frequency of the driving field, giving rise to drifts in the reference frame where $d_1 = 0$ \cite{milian_soliton_2014, parra2014third}. Thus, by choosing $d_1$ to match the rate of solitons' intrinsic drift, we reestablish synchronicity between the inhomogeneity and the soliton, forcing both to be stationary in the new reference frame. In what follows, we will refer to this reference frame as the solitons' intrinsic reference frame and, unless otherwise specified, operate in it.

In the absence of pump inhomogeneity, TCSs correspond to steady-state solutions of Eq.~(\ref{lle}) (when the equation is expressed in the solitons' intrinsic reference frame with appropriate $d_1$). This situation can change, however, in the presence of inhomogeneities. Assuming a TCS to be located at $\tau_\mathrm{cs}$, and writing without loss of generality $S(\tau) = S_0 + P(\tau)$, where $P(\tau_\mathrm{cs}) = 0$, theory predicts that the TCS will drift due to the pump inhomogeneity at a rate given by \cite{maggipinto2000cavity, scroggie2005reversible}
\begin{align} \label{speed}
    v \equiv \dv{\tau_\mathrm{cs}}{t} = - \frac{\braket{\mathrm{v}_0(\tau-\tau_\mathrm{cs})}{P(\tau)}}{\braket{\mathrm{v}_0(\tau)}{\displaystyle \dv{E_\mathrm{s}(\tau)}{\tau}}}.
\end{align}
Here, $E_\mathrm{s}(\tau)$ is the TCS solution to the homogeneous LLE when centred at $\tau_\mathrm{cs} = 0$, $\mathrm{v}_0(\tau)$ is the neutral (or Goldstone) mode corresponding to $E_\mathrm{s}(\tau)$ (associated with translation symmetry in the homogeneously driven LLE), and the inner product is defined as
\begin{align} \label{integrals}
    \braket{f(\tau)}{g(\tau)} \equiv &\int_{-\infty}^\infty f_r(\tau) g_r(\tau) \dd{\tau} \nonumber\\
    &+ \int_{-\infty}^\infty f_i(\tau) g_i(\tau) \dd{\tau},
\end{align}
where the subscripts $r$ and $i$ denote the real and imaginary parts of the (in general) complex-valued functions, respectively. Note that the minus sign in front of the right-hand side of Eq.~(\ref{speed}) corrects a misprint that is present in existing literature \cite{erkintalo2022phase}. Also note that the neutral mode $\mathrm{v}_0(\tau)$ exhibits a sign ambiguity: in what follows, we choose the sign such that the denominator in Eq.~(\ref{speed}) is positive.

Typically, the perturbation $P(\tau)$ is assumed to change slowly in comparison to the TCS profile, such that a first-order Taylor series expansion is sufficient, i.e, \mbox{$P(\tau) \approx \dv{P}{\tau} \eval_{\tau_\mathrm{cs}} (\tau - \tau_\mathrm{cs})$}. In this case, we have
\begin{align} \label{linearspeed}
    v = - N \bigg( \braket{\mathrm{v}_{0r}(\tau-\tau_\mathrm{cs})}{\tau-\tau_\mathrm{cs}} &\dv{P_r}{\tau} \eval_{\tau_\mathrm{cs}} \nonumber\\[5pt]
    + \braket{\mathrm{v}_{0i}(\tau-\tau_\mathrm{cs})}{\tau-\tau_\mathrm{cs}} &\dv{P_i}{\tau} \eval_{\tau_\mathrm{cs}}  \bigg),
\end{align}
where we have defined the (positive) normalizing factor
\begin{align}
    N \equiv \frac{1}{\braket{\mathrm{v}_0}{\displaystyle \dv{E_\mathrm{s}}{\tau}}}.
\end{align}
When the perturbation is purely real (e.g., for driving field amplitude modulation~\cite{hendry2018spontaneous}) or purely imaginary (e.g., for driving field phase modulation~\cite{jang2015temporal}), Eq.~(\ref{linearspeed}) amounts to the conventional result that the induced drift is proportional to the gradient of the perturbation at the TCS position. In contrast, our goal is to demonstrate that this linear approximation is not always valid in physically realizable circumstances and to explore the implications that arise as a result. While it would be tempting to approach this problem by simply including more Taylor series expansion terms to describe $P(\tau)$, this plan of attack is unwieldy in general, often requiring hundreds of (computationally expensive) terms before convergence is achieved. Rather, in what follows, we will consider the full overlap between the neutral mode and the inhomogeneity and use Eq.~(\ref{speed}) to gain insights on TCS motion. We achieve this by finding the steady-state TCS solutions of Eq.~\eqref{lle} with a multi-dimensional Newton-Raphson method, which then allows us to compute the neutral mode as described in~\cite{maggipinto2000cavity, erkintalo2022phase}. To confirm our calculations, we also perform direct dynamical simulations of Eq.~\eqref{lle} using a standard split-step Fourier algorithm.

\section{Results}

\subsection{Impact of rapid parameter inhomogeneities}

\begin{figure}[t]
    \centering
    \includegraphics[scale=1]{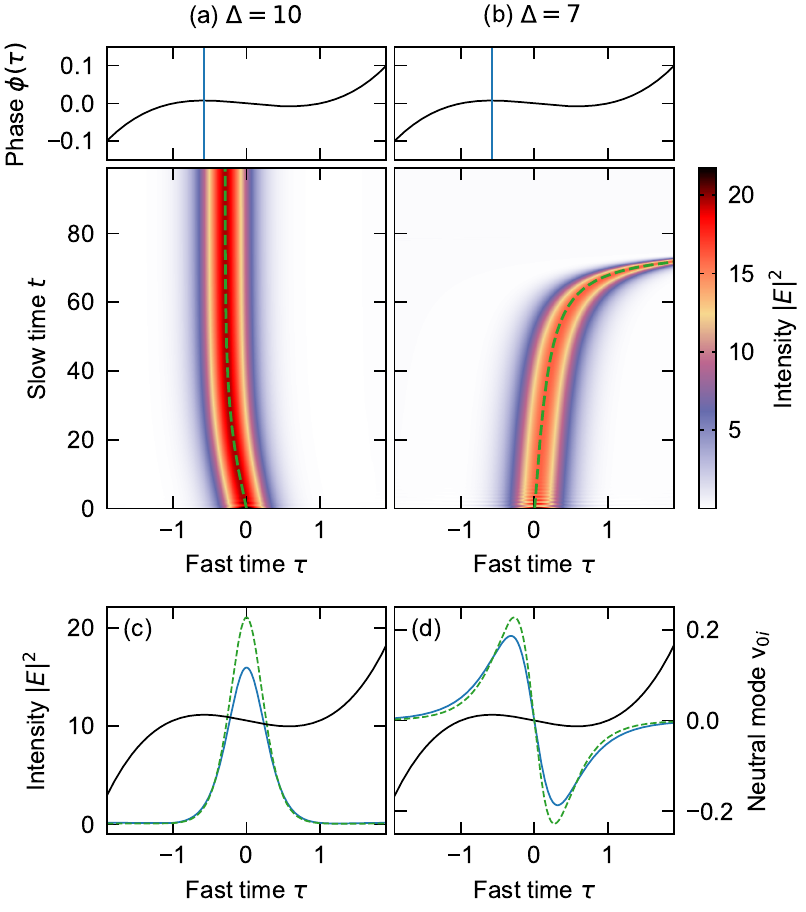}
    \caption{(a, b) Numerical simulation results, showing TCS dynamics with driving amplitude $S_0 = \sqrt{10}$ and an applied driving phase modulation $\phi(\tau)$ given by Eq.~(\ref{phase}) with $p = 0.02$ and $q = 1$ for two different detunings: (a) $\Delta = 10$; (b) $\Delta = 7$. The top panels show the cubic phase profile (with the local maximum indicated as a vertical line) while the bottom panels show the simulated soliton dynamics. Green dashed curves show soliton trajectories predicted from Eq.~(\ref{speed}). (c, d) TCS intensity profiles (c) and imaginary components of the neutral modes (d) for $\Delta = 10$ (green dashed curves) and $\Delta = 7$ (blue solid curves) compared to $\phi(\tau)$ (black, rescaled vertically for clarity).}
    \label{fig:cubic_comparison}
\end{figure}

We begin by providing an illustrative example on how conventional wisdom based on the assumption of gradient motion may fail. To this end, we ignore third-order dispersion ($d_3 = 0$) and consider a driving field that is phase modulated, $S(\tau) \equiv S_0 \exp[i \phi(\tau)]$, with the phase profile $\phi(\tau)$ being a cubic polynomial in the vicinity of the soliton:
\begin{align} \label{phase}
    \phi(\tau) \equiv p \tau (\tau + q) (\tau - q),
\end{align}
where $p$ and $q$ are positive real numbers. If $p$ is chosen to be sufficiently small, we have in the vicinity of the TCS \mbox{$S(\tau) \approx S_0 [1 + i \phi(\tau)] = S_0 + i S_0 \phi(\tau)$}, yielding a purely imaginary perturbation \mbox{$P(\tau) = i S_0 \phi(\tau)$}. In this case, the linear approximation predicts that the induced drift rate $v \approx 2 \dv{\phi}{t}$, such that a TCS initially located at $\tau_\mathrm{cs} = 0$ will drift towards the local phase maximum at \mbox{$\tau_\mathrm{M} = -\sqrt{q^2/3}$} \cite{firth1996optical, jang2015temporal}. While (approximately) true for large values of $q$, this prediction fails as $q$ gets small enough such that the phase gradient changes substantially across the TCS width. Figure~\ref{fig:cubic_comparison} shows results from numerical simulations of Eq.~(\ref{lle}) that illustrate this point. Here we consider TCS dynamics in the presence of a cubic phase profile with \mbox{$p = 0.02$} and \mbox{$q = 1$} with constant driving intensity \mbox{$S_0^2 = 10$} but for two different detunings \mbox{$\Delta = 10$} [Fig.~\ref{fig:cubic_comparison}(a)] and \mbox{$\Delta = 7$} [Fig.~\ref{fig:cubic_comparison}(b)]. Despite the solitons being associated with identical pump inhomogeneity and identical initial position \mbox{($\tau_\mathrm{cs} = 0$)}, we observe starkly different dynamics for the two detunings. For \mbox{$\Delta = 10$}, the soliton drifts up the gradient and becomes trapped near (but not at) the phase maximum, whilst for \mbox{$\Delta = 7$} the soliton drifts \emph{down} the gradient. While not shown, we note that for \mbox{$\Delta \approx 7.5$}, the soliton does not move at all in spite of the local phase gradient.

The qualitatively different behaviors observed in Figs.~\ref{fig:cubic_comparison}(a) and (b) can be understood by recalling that the width of a TCS scales as $1/\sqrt{\Delta}$ \cite{coen2013universal}, with similar scaling implied to the corresponding neutral mode $\mathrm{v}_0(\tau)$ [see Figs.~\ref{fig:cubic_comparison}(c) and (d)]. Specifically, for \mbox{$\Delta = 10$}, the neutral mode is comparatively more localized in the vicinity of \mbox{$\tau = 0$}, where the product between the neutral mode and the phase perturbation yields positive values, thus resulting in an overall positive value for the corresponding overlap integral [and hence a negative velocity, see Eq.~(\ref{speed})]. As a consequence, the soliton drifts towards the local phase maximum at $\tau_\mathrm{M}$, but ultimately halts at a position where the overlap between the (imaginary component of) the neutral mode and the phase perturbation becomes zero (this position can be seen to be slightly offset from the phase maximum). In stark contrast, because the neutral mode is less localized for \mbox{$\Delta = 7$}, the full overlap between the phase perturbation and the neutral mode yields a negative value (and hence a positive velocity). As a result, the TCS is pushed away from the local phase maximum with an accelerating drift rate in a way similar to that observed in \cite{scroggie2005reversible} for spatial cavity solitons. In both cases, we find that Eq.~(\ref{speed}) provides a good quantitative agreement with the observed soliton trajectories [dashed green curves in Fig.~\ref{fig:cubic_comparison}(a) and (b)]. Although not shown here, we remark that our simulations reveal that the TCS for \mbox{$\Delta = 7$} eventually ceases to exist because the phase perturbation varies too rapidly across it for sufficiently large $\tau_\mathrm{cs}$.

\subsection{Bichromatic driving}

The results shown in Fig.~\ref{fig:cubic_comparison} clearly illustrate that, for sufficiently rapid perturbations, TCS motion is not solely determined by the local gradient of the perturbation. While it is unlikely that direct (electronic) phase modulation can yield such rapid perturbations in practice, the use of bichromatic driving~\cite{hansson_bichromatically_2014,qureshi_soliton_2021,moille_ultra-broadband_2021,taheri2022all} can in principle yield arbitrarily fast phase modulations (along with concomitant amplitude modulations which can affect the overall dynamics). To gain more insights into such situations, we next consider a bichromatic driving field of the form \mbox{$S(\tau) = S_1 + S_2 e^{i \Omega \tau}$} and analyze how the modulation frequency $\Omega$ (the angular frequency spacing of the driving fields) qualitatively changes the soliton motion. For the sake of simplicity, we still consider the situation where third-order dispersion is negligible (\mbox{$d_3 = 0$}).

The analysis pertaining to bichromatic driving is similar to the analysis considered by earlier works in the context of localized structures in diffractive systems in the presence of sinusoidal perturbations \cite{fedorov2001effects, scroggie2005reversible}. We first define $P(\tau)$ such that it satisfies the assumption \mbox{$P(\tau_\mathrm{cs}) = 0$} by rewriting the driving field as:
\begin{align} \label{bichromaticdriving}
    S(\tau) = S_0 + S_2 ( e^{i \Omega \tau} - e^{i \Omega \tau_\mathrm{cs}} ),
\end{align}
where \mbox{$S_0 \equiv S(\tau_\mathrm{cs}) = S_1 + S_2 e^{i \Omega \tau_\mathrm{cs}}$} is the complex amplitude of the driving field at the position of the soliton. Substituting the perturbation \mbox{$P(\tau) = S_2 ( e^{i \Omega \tau} - e^{i \Omega \tau_\mathrm{cs}} )$} into Eq.~(\ref{speed}) yields
\begin{align} \label{bichromaticspeed}
    v = - N \braket{\mathrm{v}_0(\tau-\tau_\mathrm{cs})}{S_2 ( e^{i \Omega \tau} - e^{i \Omega \tau_\mathrm{cs}} )}.
\end{align}
When $d_3 = 0$, the neutral mode, $\mathrm{v}_0(\tau-\tau_\mathrm{cs})$, is an odd function. In this case, Eq.~(\ref{bichromaticspeed}) can be simplified (after some algebra) into
\begin{align} \label{fourierspeed}
    v = N S_2 \Big[ u_r(\Omega) \sin(\Omega \tau_\mathrm{cs}) - u_i(\Omega) \cos(\Omega \tau_\mathrm{cs}) \Big],
\end{align}
where \mbox{$u_r(\Omega) = \Im[\widetilde{\mathrm{v}}_{0r}(\Omega)]$} and \mbox{$u_i(\Omega) = \Im[\widetilde{\mathrm{v}}_{0i}(\Omega)]$} are the imaginary parts of the Fourier transforms of the real and imaginary parts of the neutral mode defined through:
\begin{align}
    \widetilde{\mathrm{v}}_{0x}(\Omega) \equiv \int_{-\infty}^\infty \mathrm{v}_{0x}(\tau) e^{i \Omega \tau} \dd{\tau}.
\end{align}

\begin{figure}[t]
    \centering
    \includegraphics[scale=1]{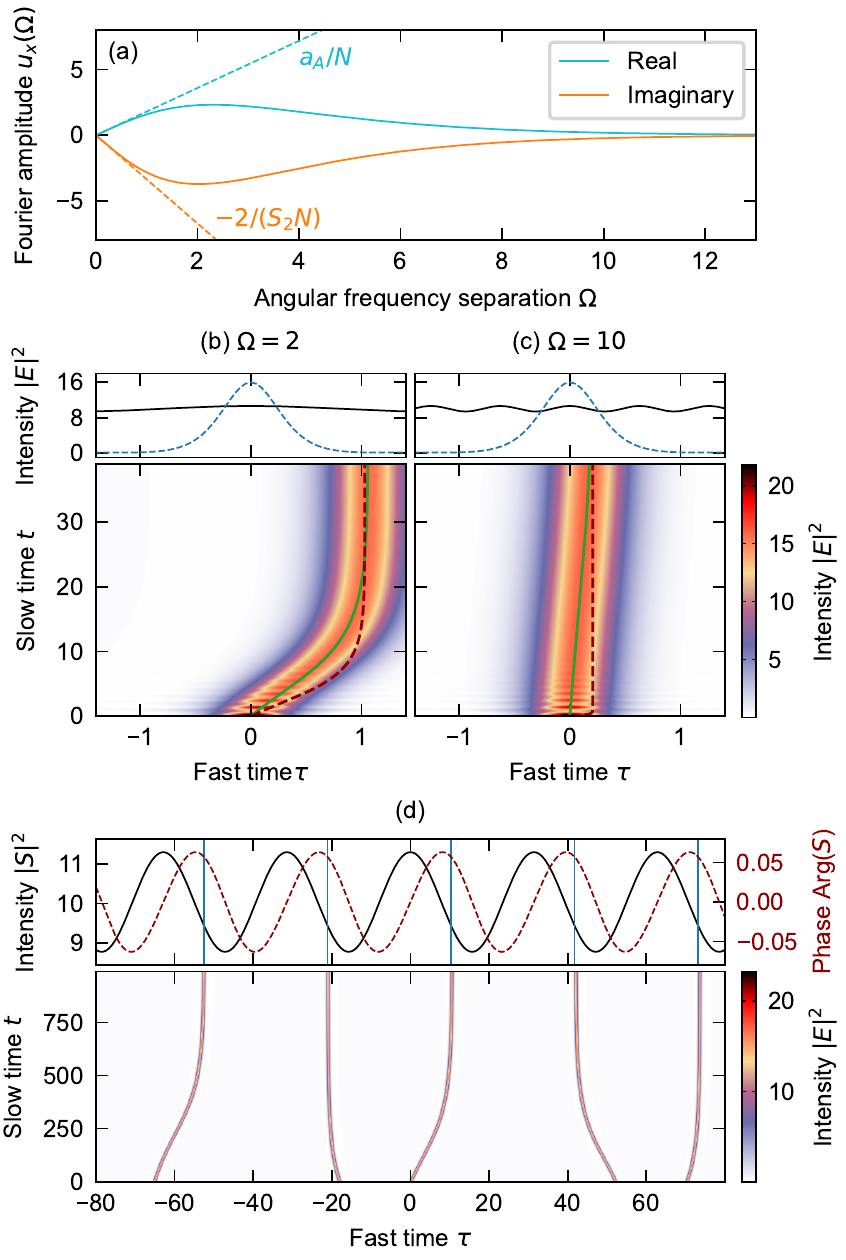}
    \caption{(a) $u_r(\Omega)$ (cyan solid line) and $u_i(\Omega)$ (orange solid line) at $\Delta = 7$ and \mbox{$|S_0| = \sqrt{10}$} compared with the corresponding linear predictions (dashed lines with denoted gradients). (b, c) Simulation results for a bichromatic driving field with $\Delta = 7$, \mbox{$S_1 = \sqrt{10}$}, \mbox{$S_2 = 0.1$}, and (b) \mbox{$\Omega = 2$} and (c) $\Omega = 10$. The top panels show the initial TCS profiles (blue dashed lines) superimposed with the intensity of the driving field $|S(\tau)|^2$ (black solid lines) on the same scale. Pseudo-color plots in the bottom panels show the simulated TCS dynamics while the dashed red and solid green curves show predictions from the linear and Fourier theories, respectively. (d) Simulation of solitons initialized within a bichromatic driving field now with $S_2 = 0.2$ and $\Omega = 0.2$. The top panel shows the driving field intensity (solid black) and phase (dashed red) with TCS trapping points predicted by Eq.~(\ref{bichromatictrapping}) shown as vertical blue lines. The bottom panel is a pseudo-color plot of soliton trajectories.}
    \label{fig:fourier_amplitudes}
\end{figure}

Equation~(\ref{fourierspeed}) shows that the TCS velocity in the presence of two driving fields separated by angular frequency $\Omega$ is dependent upon the Fourier amplitudes of the real and imaginary parts of the neutral mode evaluated at $\Omega$. To gain more insights, we plot these Fourier coefficients in Fig.~\ref{fig:fourier_amplitudes}(a) for $\Delta = 7$ and $|S_0| = \sqrt{10}$. As can be seen, the magnitudes of both coefficients grow linearly for small $\Omega$, but quickly deviate from this trend to peak at $\Omega \approx 2$ and subsequently decay to zero. The initial linear growth of the Fourier transforms is equivalent to gradient motion, as can be readily seen by using $u_x(\Omega) \propto \Omega$ in Eq.~(\ref{fourierspeed}). For $u_i(\Omega)$, the low-frequency slope is equal to $-2/(N S_2)$, as expected to maintain congruence with the familiar expression $v = 2 \dv{\phi}{\tau}$ known to hold for (purely imaginary) phase perturbations~\cite{firth1996optical, jang2015temporal, erkintalo2022phase}. In a similar manner, the slope of $u_r(\Omega)$ is equal to $a_\mathrm{A}(\Delta, |S_0|)/N$, where $a_\mathrm{A}(\Delta, |S_0|)$ is the drift coefficient that describes TCS motion under presence of pure (real) amplitude inhomogeneities \cite{hendry2018spontaneous}.

The dashed lines in Fig.~\ref{fig:fourier_amplitudes}(a) show that the linear approximation to the Fourier amplitudes is accurate until about \mbox{$\Omega \approx 1$}. However, as $\Omega$ increases, the coefficients deviate from linearity, signalling departure from gradient motion. Figures~\ref{fig:fourier_amplitudes}(b) and (c) show numerical simulations of TCS motion under bichromatic driving at two different frequency separations: $\Omega = 2$ [Fig.~\ref{fig:fourier_amplitudes}(b)] and $\Omega = 10$ [Fig.~\ref{fig:fourier_amplitudes}(c)]. In both examples, the trajectories derived from Fourier theory (green solid curves) agree very closely with the numerical observations. In Fig.~\ref{fig:fourier_amplitudes}(b), the linear prediction (red dashed curve) overestimates the initial speed by approximately a factor of two, which is consistent with the comparisons in Fig.~\ref{fig:fourier_amplitudes}(a). In Fig.~\ref{fig:fourier_amplitudes}(c), the difference is far greater: the linear prediction angles sharply towards the expected trapping point, while the Fourier prediction (and the observed trajectory) moves more slowly. This demonstrates the fact that increasing $\Omega$ will not arbitrarily increase the induced drift speeds, contradicting the linear result's assertion and highlighting how the soliton velocity decreases asymptotically as the frequency shift $\Omega$ becomes large. This latter observation can be explained physically by noting that the soliton is not able to parse the fine structure of extremely rapid oscillations and will only respond to the mean field.

Equation~(\ref{fourierspeed}) demonstrates that, for perturbations that possess both non-zero real and imaginary parts (such as bichromatic driving), the two components add linearly to yield the total TCS velocity. With knowledge of the total velocity, one can find the positions at which the TCSs will be trapped. For bichromatic driving, the fast times where \mbox{$v = 0$} are given by:
\begin{align} \label{bichromatictrapping}
    \tau_0 = \frac{1}{\Omega} \bigg[ \arctan(\frac{u_i(\Omega)}{u_r(\Omega)}) + m \pi \bigg],
\end{align}
where $m$ is an integer. Equation~(\ref{bichromatictrapping}) shows that, within each period $2\pi/\Omega$, there are in general two fixed points. However, only every other point is stable, with the stable trapping points all being on the same side of the phase (or amplitude) maxima of the driving field, and thus naturally repeating at the period of $2\pi/\Omega$. We illustrate this point in Fig.~\ref{fig:fourier_amplitudes}(d), where we show results from simulations where five TCSs are initialized at different positions along the driving field. As can been seen, all of the solitons are attracted to positions on the trailing edges of amplitude maxima, coinciding with the stable trapping positions predicted by Eq.~(\ref{bichromatictrapping}) (indicated in the top panel by vertical lines). Note that our results agree with recent findings where bichromatic driving was leveraged to realize perfectly periodic TCS sequences and hence study discrete time crystals~\cite{taheri2022all}. It is also worth noting that Figs.~\ref{fig:fourier_amplitudes}(b) and (c) give evidence that the trapping points calculated using the linear approximation [Eq.~(\ref{linearspeed})] agree closely with the full result --- though the trajectories are notably different. This is because the arctan function in Eq.~(\ref{bichromatictrapping}) deviates relatively little from its $\Omega = 0$ value over the range of inputs $u_i(\Omega)/u_r(\Omega)$, and what deviation is present is diminished by the factor of $1/\Omega$.

Before proceeding, we note that the neutral mode $\mathrm{v}_0(\tau)$ depends on the driving amplitude at the TCS position $|S_0|$. Correctly computing trajectories like those in Fig.~\ref{fig:fourier_amplitudes} will, in principle, necessitate finding the neutral mode at every driving amplitude present along the trajectory. However, if $S_2$ is small, $\mathrm{v}_0(\tau)$ (and the corresponding Fourier amplitudes) will change very little, in which case computing one $\mathrm{v}_0(\tau)$ and treating that as applicable throughout (as we did in Fig.~\ref{fig:fourier_amplitudes}) is sufficient.

\subsection{Third-Order dispersion}

So far, we have considered scenarios involving an \emph{intrinsically} rapid parameter inhomogeneity. However, even a modest inhomogeneity can become \emph{comparatively} rapid if the temporal extent of the soliton is broadened through some mechanism -- we now focus our attention to this latter scenario. Specifically, it is well known that TCSs excited close to the zero-dispersion point of the resonator develop an extended oscillatory tail due to higher-order dispersion~\cite{coen2013modeling, milian_soliton_2014, parra2014third, jang_observation_2014}. To the best of our knowledge, the impact of parameter inhomogeneities on such near-zero-dispersion TCSs~\cite{li_experimental_2020, li2021observations, anderson_zero-dispersion_2020, zhang_microresonator_2022} has not yet been investigated.

\begin{figure}[t]
    \centering
    \includegraphics[scale=1]{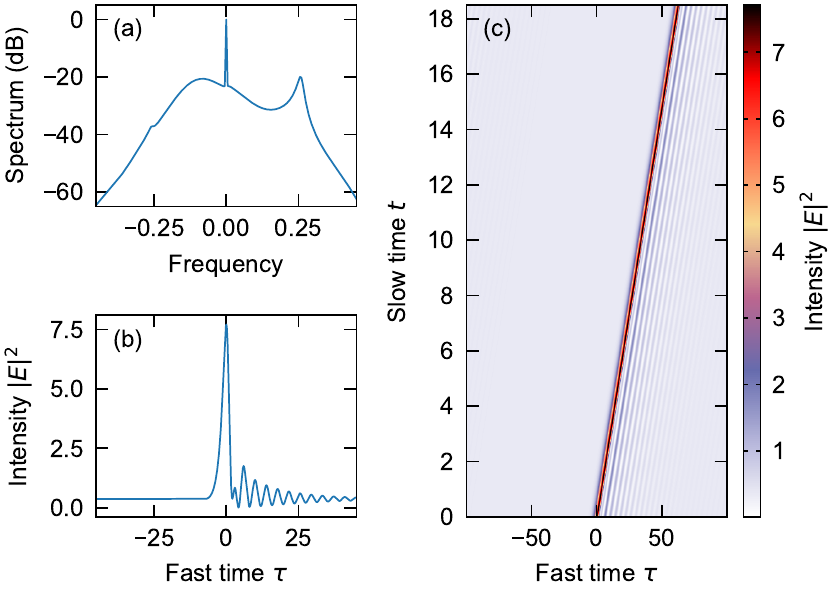}
    \caption{(a) Spectral and (b) temporal intensity profiles of a near-zero-dispersion TCS generated with $S_0 = \sqrt{10}$, $\Delta = 5.5$, and \mbox{$d_3 = 3$} in the absence of any parameter inhomogeneities. (c) A pseudo-color plot of the soliton's evolution in the reference frame moving at the group velocity at the carrier frequency of the driving field (i.e., with \mbox{$d_1 = 0$}).}
    \label{fig:d3homogeneous}
\end{figure}

Figure~\ref{fig:d3homogeneous} recalls well-known characteristics of a typical TCS in the presence of significant third-order dispersion (\mbox{$d_3 = 3$}, for other parameters, see caption) and in the absence of any parameter inhomogeneity. Due to third-order dispersion, the TCS emits a dispersive wave (DW) that appears in the intracavity spectrum as a sharp peak in the normal dispersion regime [c.f. Fig.~\ref{fig:d3homogeneous}(a)]. In the temporal domain, the dispersive wave interferes with the background on top of which the soliton sits, resulting in an extended oscillatory tail [c.f. Fig.~\ref{fig:d3homogeneous}(b)]. Moreover, the soliton (and the DW attached to it) undergoes constant temporal drift (in the pump reference frame with \mbox{$d_1 = 0$}) as shown in Fig.~\ref{fig:d3homogeneous}(c): here the soliton can be seen to drift at a rate of $v_\mathrm{d} = \dv{\tau_\mathrm{cs}}{t} \approx 3.3$.

Systematic analysis of near-zero-dispersion TCS motion in the presence of parameter inhomogeneities is significantly complicated by the solitons' intrinsic drift. This is in particular due to the rate of drift ($v_\mathrm{d}$) depending on the detuning $\Delta$ and driving amplitude $|S|$, thus forcing a different reference frame ($d_1$) to be used at all different positions when the inhomogeneity lies in either of these parameters. However, the drift rate does not depend on the phase of the driving field, allowing motions induced by phase modulations of that field to be examined in a single reference frame. We therefore focus our attention on pure phase modulations of the driving field. Moreover, in what follows, all calculations of soliton motion have been performed in the soliton's intrinsic reference frame, i.e., with the parameter $d_1$ chosen to match the soliton's drift speed under homogeneous driving. This essentially ensures that the soliton's motion is represented by a drift in the reference frame where the soliton would be stationary in the absence of inhomogeneities.

\begin{figure*}[t]
    \centering
    \includegraphics{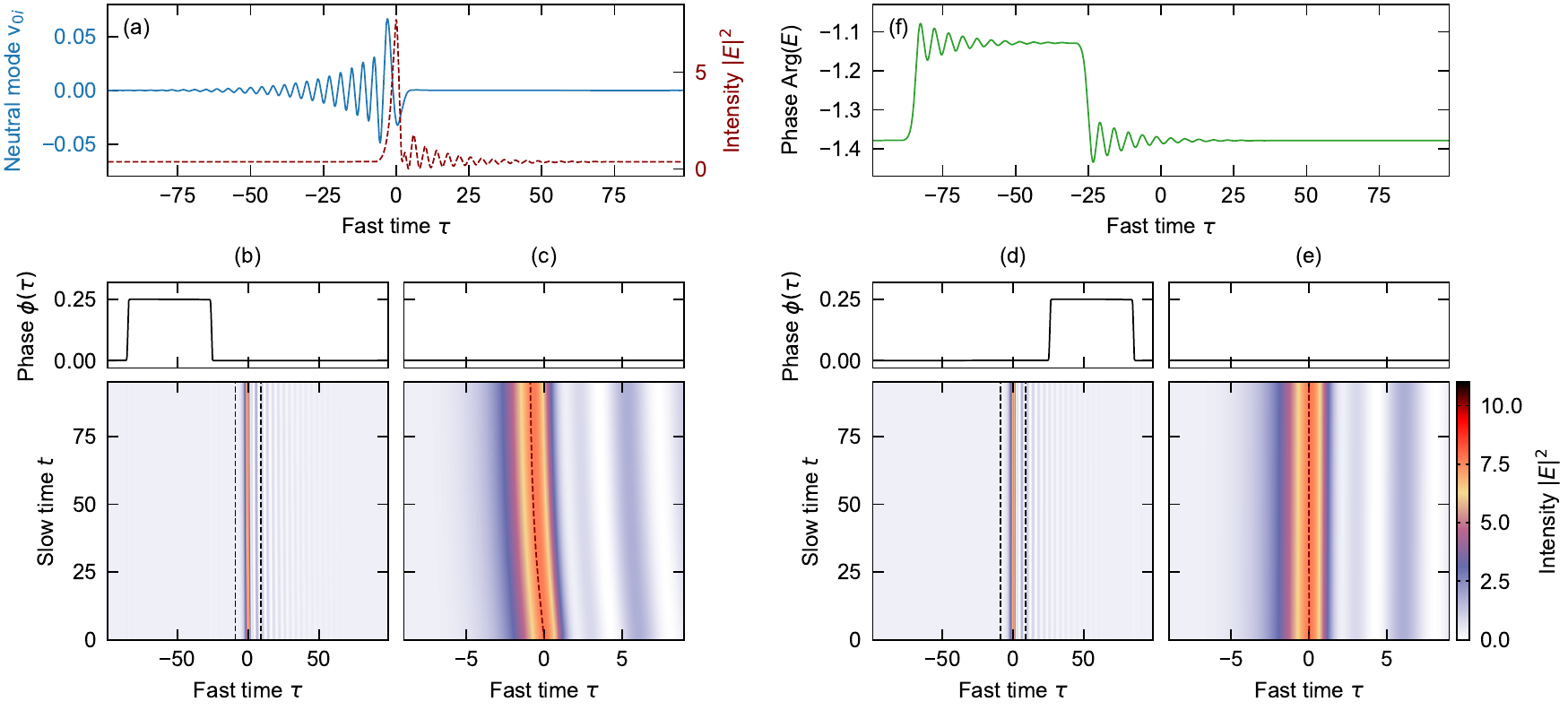}
    \caption{(a) The imaginary part of the neutral mode (blue solid curve) corresponding to a TCS with $S_0 = \sqrt{10}$, $\Delta = 5.5$, and $d_3 = 3$, whose intensity profile is shown as a red dashed curve. See Fig.~\ref{fig:d3homogeneous} for further temporal and spectral characteristics of the soliton. (b)--(e) Soliton dynamics in the presence of a semi-rectangular phase modulation that (b, c) leads and (d, e) trails the soliton. The top panels show the phase profiles of the driving field, $S = S_0\exp[i\phi(\tau)]$, while the pseudo-color plots in the bottom show dynamical simulations of soliton behaviors. (b) and (d) differ in fast time scale from (c) and (e): the former pair highlights the scale of the driving phase perturbation relative to the soliton; the latter pair focuses on the soliton core to highlight the induced motion (or lack thereof). Vertical dashed lines in (b) and (d) indicate the extent of the fast time axes in (c) and (e). (f) The steady-state intracavity field phase profile with driving parameters as in (b) and (c) but with no soliton initialized in the cavity.}
    \label{fig:neutralmode}
\end{figure*}

Figure~\ref{fig:neutralmode}(a) shows the (imaginary part of the) neutral mode (blue solid curve) associated with the near-zero-dispersion soliton (red dashed curve) considered in Fig.~\ref{fig:d3homogeneous}. As with the soliton's intensity profile, third-order dispersion manifests itself in the form of an oscillatory tail in the neutral mode. But rather surprisingly, while the DW tail trails the soliton's intensity profile, the neutral mode is associated with a tail that leads the mode. This feature is not a numerical artefact, but is representative of the TCS dynamics in the presence of third-order dispersion. To show this, we consider two driving field phase inhomogeneities that are non-zero only in regions that lead or trail the soliton. Figures~\ref{fig:neutralmode}(b)--(e) shows results from LLE simulations that consider such inhomogeneities [(c) and (e) show the same simulations as (b) and (d), respectively, but on different fast time scales]. As can be seen, an inhomogeneity that leads the soliton [Figs.~\ref{fig:neutralmode}(b) and (c)] gives rise to noticeable motion; in contrast, an inhomogeneity that trails the soliton [Figs.~\ref{fig:neutralmode}(d) and (e)] does not enact observable motion, as the neutral mode is essentially zero in the region where the inhomogeneity is non-zero. The LLE simulations shown in Fig.~\ref{fig:neutralmode}(b)--(e) clearly corroborate the fact that the neutral mode extends towards the leading edge of the soliton. As further evidence, the dashed curves in Fig.~\ref{fig:neutralmode}(c) and (e) show the soliton trajectories as predicted by the overlap integral of Eq.~(\ref{speed}), and again we observe good agreement with the simulation results.

On the one hand, the results shown in Fig.~\ref{fig:neutralmode} confirm the intuition that, by extending the soliton's temporal extent, higher-order dispersion permits parameter inhomogeneities to act upon TCSs over a broader temporal range. On the other hand, the specifics of the interaction are unexpected: the temporal domain where interactions can take place (as determined by the neutral mode) is extended in a direction that is opposite to the direction in which the soliton's intensity profile is extended (the DW tail). To understand this result, we recall that the soliton responds to the \emph{intracavity} phase profile, not the driving phase profile directly. When $d_3$ is non-zero, the intracavity phase profile develops oscillating tails --- even in the absence of a soliton --- that extend in the same direction as the oscillating tail of the soliton DW (e.g., for results in Fig.~\ref{fig:neutralmode}, the tail trails the phase inhomogeneity). This is portrayed in Fig.~\ref{fig:neutralmode}(f), which shows such an intracavity phase profile for the driving phase profile in the top panel of Fig.~\ref{fig:neutralmode}(b) but with no soliton in the resonator. When the driving phase perturbation leads the soliton, the intracavity phase tail can overlap with the soliton's core and influence its dynamics. However, a driving phase perturbation that trails the soliton has no extended tail in the negative $\tau$ direction and thus no effect on the soliton which precedes it. This directional ranged influence of phase inhomogeneities with third-order dispersion is incorporated within the neutral mode's `reversed' tail. In this context, it is worth noting that the results indicate that the soliton's DW tail does not participate in the dynamics at all, but rather, it is the overlap between the intracavity phase profile and the soliton core that dictates the drift. 

\begin{figure}[t]
    \centering
    \includegraphics[scale=1]{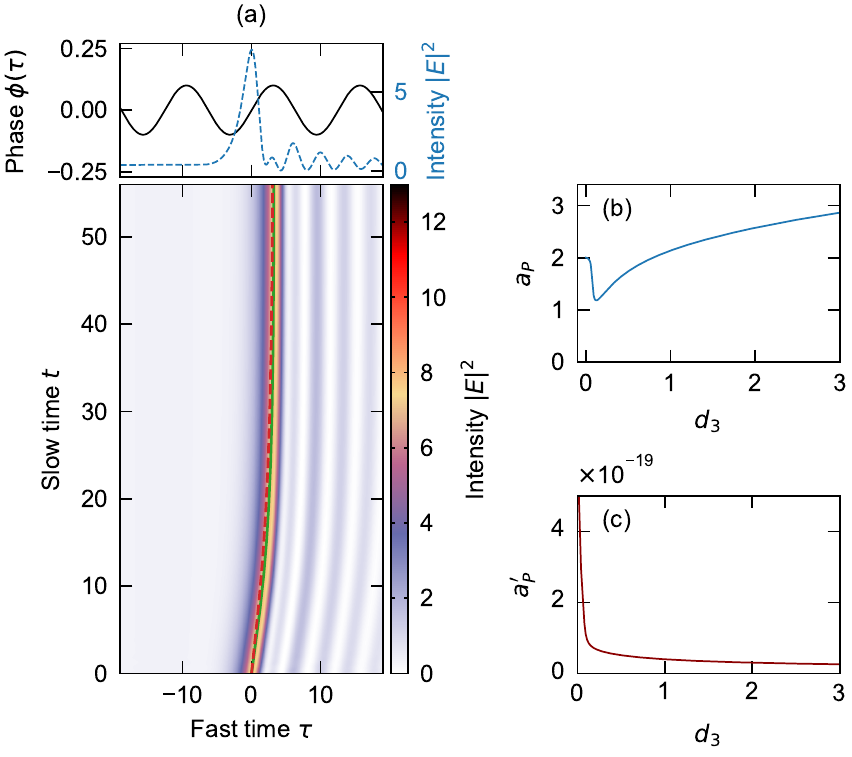}
    \caption{(a) The trajectory of a soliton with $|S| = \sqrt{10}$, \mbox{$\Delta = 5.5$,} $d_3 = 3$, and the phase profile $\phi(\tau) = 0.1 \sin(0.5 \tau)$. The top panel shows $\phi(\tau)$ (black solid curve) alongside the initial intracavity electric field intensity profile (blue dashed curve), while the bottom panel overlays the simulated dynamics with the trajectories predicted by Eq.~(\ref{speed}) (green solid curve) and its linear approximation Eq.~(\ref{linearspeed}) (red dashed curve). (b, c) The proportionality between soliton speeds and phase gradients (b) without and (c) with compensation for the normalization.}
    \label{fig:d3inhomogeneous}
\end{figure}

The extended domain through which inhomogeneities can act upon TCSs seemingly implies that the range of perturbations that satisfy the linear approximation [Eq.~(\ref{linearspeed})] is substantially constrained. However, because the tail of the neutral mode oscillates around zero [see Fig.~\ref{fig:neutralmode}(a)], we generically find that this tail contributes comparatively little to the overall soliton velocity [assuming that the perturbation does not change rapidly with respect to the oscillations, as was the case in Figs.~\ref{fig:neutralmode}(b)--(e)]. As such, the induced velocity is almost entirely determined by the perturbation near the soliton core. Despite higher-order dispersion extending the soliton's temporal profile through the DW tail, the condition for a perturbation to be sufficiently narrow for the linear approximation to be valid remains approximately the same as in the absence of higher-order dispersion. We demonstrate this point in Fig.~\ref{fig:d3inhomogeneous}(a), where we subject a soliton to a sinusoidal phase perturbation that is narrow compared to the full soliton structure but broad compared to the core of the soliton (see the top panel). The bottom panel shows the soliton's trajectory under this modulation along with the predictions of Eq.~(\ref{speed}) (green solid curve) and its linear approximation Eq.~(\ref{linearspeed}) (red dashed curve). As can be seen, the linear approximation agrees closely with both the full prediction and the simulated soliton trajectory. These results support the claim that the linear approximation can work well despite $P(\tau)$ changing nonlinearly across the entire footprint of the soliton that includes the DW tail.

Knowing that the linear approximation of soliton motion remains broadly appropriate despite an extended DW tail allows us to analyze how higher-order dispersion influences the soliton drift velocities. Considering phase modulated driving fields specifically, we obtain from Eq.~(\ref{linearspeed}) that, in the linear approximation, \mbox{$v = a_\mathrm{P} \dv{\phi}{\tau}$} where
\begin{align}
    a_\mathrm{P} = - \frac{N}{S_0} \braket{\mathrm{v}_{0i}(\tau-\tau_\mathrm{cs})}{\tau-\tau_\mathrm{cs}}.
\end{align}
Figure~\ref{fig:d3inhomogeneous}(b) shows the phase modulation drift coefficient $a_\mathrm{P}$ as a function of the third-order dispersion coefficient $d_3$ (for $\Delta = 5.5$ and $|S| = \sqrt{10}$). As expected, $a_\mathrm{P} = 2$ in the absence of third-order dispersion ($d_3 = 0$). However, as $d_3$ increases, the drift coefficient initially decreases, but quickly starts to increase for $d_3 > 0.12$.

The results shown in Fig.~\ref{fig:d3inhomogeneous}(b) suggest that the third-order dispersion coefficient $d_3$ can increase the magnitude of the soliton drift velocity. While this is true in the context of the normalized model described by Eq.~(\ref{lle}), it is important to consider the physical results in dimensional units. To this end, we first recall that Eq.~(\ref{lle}) is normalized such that~\cite{leo2010temporal}
\begin{align}
    d_3 \equiv \sqrt{\frac{2 \alpha}{9 L}} \frac{\beta_3}{|\beta_2|^{\frac{3}{2}}}
\end{align}
where $\alpha$ is half the total cavity loss, $L$ is the cavity roundtrip length, and $\beta_2$ and $\beta_3$ are the unnormalized second- and third-order dispersion coefficients, respectively. While $d_3$ can theoretically be increased by increasing the third-order dispersion coefficient $\beta_3$, in practice it is much more straightforward to decrease the second-order dispersion coefficient $|\beta_2|$ (by pumping the resonator closer to the zero-dispersion point). However, decreasing $\beta_2$ also decreases the soliton drift velocity, which in dimensional units reads
\begin{align}
    v' = \dv{\tau_\mathrm{cs}'}{t'} = \frac{L}{2 t_\mathrm{R}} |\beta_2| a_\mathrm{P} \dv{\phi}{\tau'} \equiv a_\mathrm{P}' \dv{\phi}{\tau'},
\end{align}
where $v'$, $\tau_\mathrm{cs}'$, and $t'$ are the dimensional equivalents of the unprimed versions and $t_\mathrm{R}$ is the roundtrip time. We thus see that decreasing $|\beta_2|$ directly decreases the amount of drift induced by a given (phase) inhomogeneity.

To illustrate how the physical drift velocity in dimensional units varies with increasing $d_3$, we consider the resonator used in experiments reported in ref. \cite{li2021observations}. The resonator has a roundtrip length of 5~m, a roundtrip time of 24~ns, a finesse of 400, a zero-dispersion wavelength (ZDW) of $1564.5$ nm and a third-order dispersion coefficient of $0.13$~ps$^{3}$km$^{-1}$ at the ZDW. We consider a range of pump wavelengths around the ZDW, and for each wavelength we compute the normalized third-order dispersion coefficient $d_3$ and the proportionality coefficient $a_\mathrm{P}'$. Results are shown in Fig.~\ref{fig:d3inhomogeneous}(c). As can clearly be seen, in dimensional units, the proportionality coefficient rapidly decreases with increasing $d_3$. These results thus suggest that, in real physical systems, the capability of phase modulation (or other form of inhomogeneities) to control TCS positioning and motion is significantly diminished when operating under conditions where higher-order dispersion plays a significant role (i.e., under conditions of low group-velocity dispersion $|\beta_2|$). This is physically reasonable, since the instantaneous frequency shift imparted by a phase gradient will shift the soliton's group velocity very little near-zero-dispersion. Counteracting this mechanism would require the normalized drift coefficients to increase significantly as a function of $d_3$, which does not happen [see Fig.~\ref{fig:d3inhomogeneous}(b)].

\section{Conclusions}

In summary, we have reported on a theoretical and numerical study that explores how temporal Kerr cavity solitons behave in the presence of parameter inhomogeneities that vary rapidly across the solitons' extent. We have shown how the conventional linear approximation used to examine soliton motion can fail and lead to qualitatively and quantitatively inaccurate predictions. Moreover, we have shown that the motion of TCSs in the presence of bichromatic driving fields is dictated by the Fourier transforms of the solitons' neutral mode, and we have unveiled surprising new features of TCSs in the presence of higher-order dispersion. In addition to further elucidating the dynamics of temporal Kerr cavity solitons in the presence of parameter inhomogeneities, our results could have practical implications for systems utilising bichromatic pumping or for systems in which higher-order dispersion plays an important role.

\section*{Acknowledgements}

The authors acknowledge financial support from the Marsden Fund and the Rutherford Discovery Fellowships of the Royal Society of New Zealand.

\bibliography{references}

\end{document}